\documentclass{aa}

\usepackage{psfig}
\usepackage{graphicx}
\usepackage{natbib}
\usepackage[english]{babel}  
\usepackage{txfonts}
\def\gsim{\;\lower4pt\hbox{${\buildrel\displaystyle >\over\sim}$}\,}
\def\lsim{\;\lower4pt\hbox{${\buildrel\displaystyle <\over\sim}$}\,}

\begin{document}

\title{The X-ray emission of the supernova remnant W49B observed with 
\emph{XMM-Newton}}

\author{M. Miceli\inst{1,2,3} \and A. Decourchelle\inst{1} \and 
J. Ballet\inst{1} \and F. Bocchino\inst{3} \and J.P. Hughes \inst{4} \and U. Hwang \inst{5,6} \and 
R. Petre \inst{6}}
\offprints{M. Miceli,\\ \email{miceli@astropa.unipa.it}}

\institute{DSM/DAPNIA/Service d'Astrophysique, AIM-UMR 7158, CEA Saclay, 91191 
Gif-sur-Yvette Cedex, France
\and Dipartimento di Scienze Fisiche ed Astronomiche, Sezione di 
Astronomia, Universit{\`a} di Palermo, Piazza del Parlamento 1, 90134 Palermo, 
Italy 
\and INAF - Osservatorio Astronomico di Palermo, Piazza del Parlamento 1, 
90134 Palermo, Italy 
\and Department of Physics and Astronomy, Rutgers University, 136 
Frelinghuysen Road, Piscataway, NJ 08854-8109, USA
\and Center for Astrophysical Sciences, The Johns Hopkins 
University, 3400 Charles St, Baltimore MD 21218.
\and Laboratory for High Energy Astrophysics, Goddard Space Flight Center, 
Greenbelt, MD 20771, USA
}

\date{Received, accepted}

\authorrunning{M. Miceli et al.}
\titlerunning{X-ray observation of W49B}

\abstract{In the framework of the study of supernova remnants and their complex interaction with the interstellar medium, we report on an \emph{XMM-Newton} EPIC observation of the Galactic
supernova remnant W49B.}{We investigate the spatial distribution of the chemical and physical properties of the plasma, so as to get important constraints on the physical scenario, on the dynamics of the supernova explosion, and on the interaction of the supernova remnant with the ambient interstellar clouds.}{We present line images, equivalent width
maps, and a spatially resolved spectral analysis
of a set of homogeneous regions.}{The X-ray spectrum of W49B is characterized by strong K emission lines from Si, S, Ar, Ca and Fe. In all the regions studied, the X-ray spectrum is dominated by the
emission from the ejecta and there is no indication of radial stratification of
the elements. A high overabundance of Ni ($Ni/Ni_{\odot}=10^{+2}_{-1}$) is required in the bright central region and the previous detection of Cr and Mn line emission is confirmed. Spectra are well described by two thermal components in collisional ionization equilibrium.  We observe spatial variations in
the temperature, with the highest temperature found in the east and
the lowest in the west.}{Our results support a scenario in which the
X-ray emission comes from ejecta interacting with a dense belt of
ambient material, but another possibility is that the remnant is the
result of an asymmetric bipolar explosion with the eastern jet being
hotter and more Fe-rich than the western jet.  The eastern jet is
confined by interaction with ambient molecular clouds.  Comparison of
the observed abundances with yields for hypernova and supernova
nucleosynthesis does not directly support the association of W49B with
a $\gamma$-ray burst, although it remains possible.
} 

\maketitle

\section{Introduction}
\label{Introduction}

Supernova remnants (SNRs) are powerful sources of heavy elements and
energy in the interstellar medium (ISM). The supernova supersonic
material (ejecta) is expelled into the ambient medium and produces a
shock front that heats and ionizes the ISM. As this shock expands into
the ISM, a reverse shock associated with the deceleration of the
ejecta propagates inward (in the reference frame of the expanding
ejecta), thus heating the expelled material to X-ray emitting
temperatures. The resulting X-ray emission provides important
information about the emitting conditions and the chemical composition
of the ejecta.

The galactic supernova remnant W49B is one of the brightest
ejecta-dominated SNRs observed in X-rays. It has a center-filled
morphology in X-rays and a shell-like morphology in the radio
band. According to \citet{rp98}, W49B is similar to mixed-morphology
supernova remnants, but its classification is uncertain.  W49B is one
of the 10 galactic sources with the highest surface brightness at 1
GHz (\citealt{mr94}). The diameter of the radio shell is $\sim4'$ and
the distance to the remnant, estimated on the basis of HI absorption
(\citealt{rgm72}), is $\sim8$ kpc (considering the corrections by
\citealt{mr94}).

In the X-ray band, W49B was first detected with the \emph{Einstein
Observatory} (\citealt{ptb84}). The discovery of intense Fe XXV K line
emission by \emph{EXOSAT} (\citealt{spj85}) indicated that the X-ray
emission of W49B is thermal and dominated by the ejecta.  This result
has been further confirmed by analysis of the \emph{ASCA} spectrum
(\citealt{fti95}, \citealt{hph00}).  Ejecta-dominated remnants are
expected to be relatively young, and while the age of W49B has not yet
been determined accurately, published age estimates are in the range
of $\sim1000-4000$ yr (\citealt{ptb84}, \citealt{spj85},
\citealt{hph00}).  The images of Si, S and Fe K emission lines
obtained with $ASCA$ indicate that the iron emission is confined to
the inner part of the remnant, while silicon and sulfur emission
appears to originate from the outer regions. This result has been
interpreted as evidence for a stratified distribution of the
elements (\citealt{fti95}). The analysis of the same \emph{ASCA} data
carried out by \citet{hph00} showed that broadband modeling of the
remnant's global spectrum requires two thermal components and
significant overabundances of Si, S, Ar, Ca, Fe with respect to the
solar values. Recently, the comparison between data obtained in X-rays
with the \emph{Chandra} satellite and in the infrared with the Palomar
observatory have revealed new details of the morphology of W49B
(\citealt{krr04}). In particular, the presence of an elongated
structure in the Fe K emission (interpreted by the authors as a jet),
and the nondetection of a neutron star associated with the remnant,
have been considered to make W49B consistent with theoretical
predictions for a collapsar (e.g., \citealt{mw99}) and thus a
candidate for a gamma-ray burst remnant.

Here we present the analysis of an \emph{XMM-Newton} EPIC observation
of W49B.  The high sensitivity of \emph{XMM-Newton} provides deep
insights into the spatial distribution of the elements and the
physical conditions in the remnant which lead to important constraints
on the physical scenario.  The paper is organized as follows:
Sect. \ref{The Data} presents the EPIC data and describes the data
processing procedure; Sect. \ref{X-ray emission} presents the X-ray
results in terms of the global spectrum (\ref{Global spectrum}),
imaging (\ref{line images}) and spatially resolved spectral analysis
(\ref{spectral analysis}). Sect. \ref{Discussion} discusses the
results in the context of observations at other wavelengths and SN
models.  Finally, we draw our conclusions in Sect. \ref{Summary and
conclusions}.

\section{Data processing}
\label{The Data}

The \emph{XMM-Newton} data presented here consist of two observations
Observation ID 0084100401 and 0084100501 (PI A.  Decourchelle)
performed on 3 April 2004 and 5 April 2004 with the EPIC MOS
(\citealt{taa01}) and EPIC pn (\citealt{sbd01}) cameras, as summarized
in Table \ref{tab:data} . Both observations have pointing coordinates
$\alpha$ $(2000)=19^{h}11^{m}13^{s}$ and $\delta$
$(2000)=9^\circ7'1''$. A third data set 0084100601 (with the same
pointing coordinates) is heavily contaminated by proton flares and was
not used for our analysis.
\begin{center}
\begin{table}[htb!]
\begin{center}
\caption{Relevant information about the data.}
\begin{tabular}{lccccc} 
\hline\hline
OBS$\_$ID & CAMERA  &  $t_{exp}$ (ks)$^{*}$  &        Mode       &  Filter  \\ \hline
0084100401 &  MOS1   &       18.6/15.8        &    Large Wind.   &  medium  \\ 
0084100401 &  MOS2   &       18.7/16.0        &    Large Wind.   &  medium  \\
0084100401 &   pn    &       17.0/12.9        &    Full Frame     &  medium  \\
0084100501 &  MOS1   &       18.7/16.8	      &    Large Wind.   &  medium  \\ 
0084100501 &  MOS2   &       18.7/16.3        &    Large Wind.   &  medium  \\
0084100501 &   pn    &       17.0/14.6        &    Full Frame	  &  medium  \\
\hline\hline
\multicolumn{5}{l}{\footnotesize{* Unscreened/Screened exposure time.}} \\
\label{tab:data}
\end{tabular}
\end{center}
\end{table}
\end{center}

The data were processed using the Science Analysis System (SAS
V6.1.0). To create light curves, images and spectra, we selected
events with PATTERN$\le12$ for the MOS cameras, PATTERN$\le4$ for the
pn camera, and FLAG$=0$ for both. To eliminate the contamination by
soft proton flares, we screened the data by applying a count-rate
limit on the light curves at high energies ($10-12$ keV for MOS and
$12-14$ keV for the pn camera). With a $\delta t=100$ s binning for
the light curves, we set the count-rate limit at 18 counts per bin for
the MOS cameras, and 22 counts per bin for the pn camera. The exposure
times are given in Table \ref{tab:data}.  For each camera we merged
the screened event files of the two observations using the task
$MERGE$.

All the images presented here are superpositions of the MOS1, MOS2 and
pn images (obtained using the $EMOSAIC$ task) and are
background-subtracted, vignetting corrected and adaptively smoothed.
Despite being relatively small in extent compared to the
\emph{XMM-Newton} field of view, W49B has a high X-ray brightness and
contaminates the local background. For this reason, we used the high
signal-to-noise background event files E1$\_$fm0000$\_$M1.fits,
E1$\_$00fm00$\_$M2.fits, E1$\_$0000fm$\_$PN.fits (obtained using the
medium filter) described in detail in \citet{rp03}. The exposure and
vignetting correction were performed by dividing the superposed count
images by the corresponding superposed exposure maps (produced using
the task $EEXPMAP$). The pn exposure maps were scaled by the ratio of
the pn/MOS effective areas to make MOS-equivalent superposed count
rate images. We smoothed the images adaptively to a signal-to-noise
ratio of 10 using the task $ASMOOTH$.  The background-subtracted count
images and the background- and continuum-subtracted line images (Sect.
\ref{line images}) were smoothed while accounting for their variances.
The resulting template was then applied to smooth the associated
exposure maps before performing the image division. The continuum
under the lines was estimated by scaling a continuum band adjacent to
the line emission using the continuum spectral slope derived from the
global remnant spectrum. The energy ranges chosen for the lines and
the continua are listed in Table \ref{tab:lines}. In all the line
bands, the background is $\la0.5\%$ of the total signal, except for the
$7.65-8.0$ keV band where it is $\sim 3.6 \%$ of the total signal. In
the continuum band $4.4-6.2$ keV, the background amounts to $\sim 1.3
\%$.  In all the images North is up and East is on the left.

For spectral analysis, we processed the event files with the
$EVIGWEIGHT$ task (\citealt{ana01}) to correct for vignetting
effects. For the background subtraction we used the same background
event files used for the images.  Spectra were rebinned to achieve a
signal-to-noise ratio per bin $>5\sigma$, and the spectral fits
were performed simultaneously on the MOS1, MOS2 and pn spectra in the
$1-9$ keV band using XSPEC (\citealt{arn96}). The chosen on-axis
response files are m1,2$\_$medv9q20t5r6$\_$all$\_$15.rsp for MOS1,2
spectra and epn$\_$ff20$\_$sdY9$\_$medium.rsp for pn spectra. All the
reported errors are at 90\% confidence, as calculated according to
\citet{lmb76}.

\section{X-ray emission}
\label{X-ray emission}

The broadband ($1-9$ keV) X-ray morphology of W49B obtained with the
EPIC cameras is shown in Fig. \ref{fig:1-9-radio} (left panel). For
comparison, the VLA radio image obtained at 327 MHz by \citet{llk01}
is also presented in Fig. \ref{fig:1-9-radio} (right panel).  The
X-ray surface brightness of W49B has its maximum at the center of the
remnant, and shows an elongated barrel-shaped structure extending for
$\sim~5$~pc along its east-west axis (labelled $b$, see
Fig.~\ref{fig:1-9-radio}) .

At the eastern end of the remnant there is a bright elongated X-ray
structure (labelled $a$, see Fig. \ref{fig:1-9-radio}), running
perpendicular to the X-ray barrel and parallel to the radio shell
(labelled $\alpha$, see Fig.~\ref{fig:1-9-radio}). The eastern side of
W49B is bounded by a large molecular cloud detected in the 2.12 $\mu$m
molecular hydrogen narrow band image of W49B
(Fig. \ref{fig:chandraIR}).  The X-ray barrel $b$ lies between two
radio structures ($\beta$ and $\gamma$) which are parallel to the
radio shell ($\alpha$) and to the H$_{2}$ emission. The positions of
these structures ($\beta$ and $\gamma$) are almost coincident with two
coaxial ring-like structures observed in the 1.64 $\mu$m [Fe II] image
of W49B (Fig. \ref{fig:chandraIR}). The X-ray structures $a$ (east)
and $c$ (west) lie outside these radio structures. No intense X-ray
emission has been detected in the south-western region of the shell,
where copious radio and [Fe II] emission are observed ($\delta$, see
Fig.~\ref{fig:1-9-radio}). According to \citet{llk01}, there is
evidence for stronger interstellar absorption in this direction.  It
is reasonable to conclude, however, that no hard X-ray emission is
present in this region.

\subsection{Global spectrum}
\label{Global spectrum}

The impressive global EPIC spectrum of W49B in the $1-9$ keV energy
band is shown in Fig. \ref{fig:specTOT}. The spectrum is dominated by
intense emission lines of He-like and H-like ions of Si, S, Ar and Ca
and by a huge Fe K emission line blend.  A phenomenological modelling
of the global spectrum (2 bremsstrahlung for the continuum $+$ narrow
gaussians for the lines) allows us to derive the line energies
reported in Fig.~\ref{fig:specTOT}.  The emission feature observed at
$\sim~7.8$~keV corresponds to a possible blending of Fe XXV and Ni
XXVII K emission lines and is discussed in sect.~\ref{Ni+Cr}.  The
continuum above 2.3 keV is well described by a thermal bremsstrahlung
model with $kT=1.69$ keV and $N_{H}=5.3\times~10^{22}$~cm$^{-2}$; this
model has been used to estimate the continuum under the lines given in
Table~\ref{tab:lines}.  Unlike the $ASCA$ spectrum \citep{hph00}, the
\emph{XMM-Newton} global spectrum cannot be correctly fitted by a
two-component thermal model (reduced $\chi^{2}\sim2.4$). This
indicates that the emission originates from physically inhomogeneous
regions. Spatially resolved spectral analysis of homogeneous regions
is required to correctly reproduce the spectra with 2-component
models, as discussed in Sect.~\ref{spectral analysis}.

In the analysis of the $ASCA$ global spectrum of W49B, \citet{hph00}
found evidence for the presence of two new emission line features (at
$\sim 5.7$ keV and $\sim 6.2$ keV) that are not included in the
currently available atomic database for X-ray spectral modeling. These
were attributed to the He-like Cr transitions (forbidden line at 5.655
keV; resonance line at 5.682 keV) and the He-like Mn transitions
(forbidden line at 6.151 keV; resonance line at 6.181 keV). Our
\emph{XMM-Newton} data confirm this detection. Modelling the global
spectrum of the remnant in the $4.4-6.4$ keV band with a thermal
bremsstrahlung continuum, we found a clear excess in the residuals
corresponding to these transitions.  These lines can be reproduced by
adding two gaussian components to the bremsstrahlung model to
determine the line energy and flux.  We found, respectively for Cr and
Mn, line centroid energy $E_{Cr}=5.66\pm0.01$~keV and flux
$F_{Cr}=2.5\pm0.4\times10^{-5}$~photons~cm$^{-2}$~s$^{-1}$,
$E_{Mn}=6.19^{+0.01}_{-0.04}$~keV and
$F_{Mn}=1.0\pm0.3\times10^{-5}$~photons~cm$^{-2}$~s$^{-1}$, in quite
good agreement with the values reported by \citet{hph00}:
$E_{Cr}=5.69^{+0.02}_{-0.03}$~keV and
$F_{Cr}=3.0^{+0.8}_{-1.1}\times10^{-5}$~photons~cm$^{-2}$~s$^{-1}$;
$E_{Mn}=6.17\pm0.05$ keV and
$F_{Mn}=1.3^{+1.4}_{-0.6}\times10^{-5}$~photons~cm$^{-2}$~s$^{-1}$.
The addition of the Cr gaussian leads to a reduction of $\chi^{2}$,
$\delta\chi^{2}=111$ for 616 d. o. f.; adding the Mn gaussian gives an
additional $\delta'\chi^{2}=28$ (to be compared with the values
obtained with the $ASCA$ spectrum: $\delta\chi^{2}=20$ for 61
d. o. f. and $\delta'\chi^{2}=10$).  We found no indication for the
presence of the Ti He$\alpha$ line (at $4.97$ keV), obtaining an upper
limit for the flux $F_{Ti}<3.6\times
10^{-6}$~photons~cm$^{-2}$~s$^{-1}$, that is more stringent that the
one reported from ASCA ($F_{Ti}<2\times
10^{-5}$~photons~cm$^{-2}$~s$^{-1}$, \citealt{hph00}).  The flux of
the Cr and Mn lines contributes $\sim 3\%$ of the total flux in the
$4.4-6.4$ band.


\subsection{Imaging}
\label{line images}

The spatially resolved spectral study of the emission in the different
lines provides information about the spatial distribution of the
synthesized elements and the chemical composition of the X-ray
emitting plasma. On the basis of low angular resolution $ASCA$ images
of W49B, \citet{fti95} argued that the sulfur emission, unlike the
iron emission, was associated with the outer envelope of the remnant
and not related to the bright $4-6$ keV continuum emission from the
center.

To investigate both this claim and the suggested radial stratification
of the elements, we produced background- and continuum-subtracted line
images of S, Ar, Ca, Fe and Ni lines (see Table~\ref{tab:lines}).  Si
emission line images, shown in Fig. \ref{fig:Si}, are discussed in
Sect.~\ref{spec results}.  The S H-like line image is very similar to
the S He-like line image and therefore is not shown in the figure.
All the line images present a bright centrally barrel-shaped
morphology similar to the $4.4-6.2$~keV continuum emission, as shown
in Fig.~\ref{fig:lines}. Overall, there are no significant differences
between the morphologies in the Si, S, Ar, Ca and Fe lines (although
some detailed structures may differ as we will discuss later). This
result is not in agreement with the line imaging results obtained for
the lower-Z elements with $ASCA$ data (see Fig.~4 in \citealt{fti95}).

\begin{center}
\begin{table*}[htb!]
\begin{center}
\caption{Parameters used for the production of the line images of Fig. \ref{fig:lines} and the equivalent width maps of Fig. \ref{fig:EQW}.}
\begin{tabular}{lccc} 
\hline\hline
Emission line      & Line energy band (keV) & Continuum energy band (keV) \\ \hline
 S XV K$\alpha$   &      $2.35-2.53$       &       $2.75-3.0$            \\ 
 S XVI K$\alpha$   &      $2.58-2.7$        &        $2.75-3.0$           \\
Ar XVII K$\alpha$  &      $3.05-3.2$        &        $2.75-3.0$           \\ 
 Ca XIX K$\alpha$  &      $3.8-4.0$         &        $3.45-3.7$           \\ 
 Fe XXV K$\alpha$  &      $6.45-6.9$        &        $4.4-6.2$            \\ 
Fe XXV$+$Ni XXVII K&      $7.65-8.0$        &        $7.05-7.5$           \\
\hline\hline
\label{tab:lines}
\end{tabular}
\end{center}
\end{table*}
\end{center}

A detailed analysis of the Fe K line image in Fig. \ref{fig:lines}
(lower left panel) shows that the iron emission differs from that
lower-Z elements in the western part of the remnant. In particular,
regions 6 and 7 do not have significant Fe line emission, while they
do have significant emission from Si, S, Ar and Ca.  The lower
statistic Fe XXV$+$Ni XXVII image of Fig.~\ref{fig:lines} (lower right
panel) presents a morphology that is fully consistent with that of the
Fe K image.

To disentangle higher element abundances in the bright line image
structures from higher emission measure (which might be caused by
either higher density or higher emission volume), we produced
equivalent width maps as shown in Fig.~\ref{fig:EQW}.  These depend
linearly on the abundances, but are also affected by the temperature,
column density and ionization age.  These images have been constructed
by dividing the background- and continuum-subtracted line images by
the corresponding underlying continuum.  Before the division, the
estimated continuum map was adaptively smoothed using the same
template used for the background- and continuum-subtracted line
images. As explained in Sect.~\ref{The Data}, the underlying continuum
is estimated from continuum regions adjacent to the line emission, and
assumes that the continuum has a uniform spectral slope over the
entire SNR.  As shown in Fig. \ref{fig:EQW}, the S map is clearly
different from the other maps, but the equivalent width images of Ar,
Ca, Fe and Fe$+$Ni all show similar features. In particular, all these
maps show higher equivalent widths in the central and eastern parts of
the remnant than in the western part, with maxima corresponding
approximately to regions $1$ and $3$.  This is particularly true for
iron.  In the Ar map there is an indication of non-negligible
equivalent widths in the western regions 6 and 7.

As the S H-like and He-like line images are very similar, we used the
combined $2.35-2.7$ keV band to improve the statistics of the sulfur
equivalent width map as shown in Fig.~\ref{fig:EQW} (upper left
panel). The map appears patchy and shows no clear structures. Given
the relatively low line energies, the map may be affected by local
variations of $N_{H}$, as we shall see in Sect.~\ref{spec results}.

In conclusion, the line emission maps of all the observed elements
show a centrally elongated bright structure in good agreement with the
morphology of the continuum.  The equivalent width images indicate
that Ar, Ca and Fe may be more abundant in the central and eastern
parts of the remnant than in the west.  The high values of the
equivalent width are therefore not restricted to the barrel-like
structure, but extend to the eastern side of the remnant, as
illustrated in Fig.~\ref{fig:EQW} (lower right panel), where we show
the contours of the iron line image superimposed on the iron
equivalent width map.  A clear drop-off of the equivalent width is
observed towards the western side.  This effect is particularly strong
for iron, and less pronounced for the lower-Z elements which show
significant equivalent widths in the west.  While we thus find an
East/West abundance anisotropy, we find no clear evidence for radial
stratification of the elements.

\subsection{Spatially resolved spectral analysis}
\label{spectral analysis}

To quantify the equivalent width map results for the element
distributions, we performed a spatially resolved spectral analysis
that allowed us to study in detail the physical and chemical
characteristics of the plasma in different parts of the remnant. In
particular, we analyzed the spectra extracted from seven fairly
homogeneous regions shown in Fig.~\ref{fig:lines} (lower left panel)
and in Fig.~\ref{fig:EQW}. While this set of regions does not cover
the entire remnant, it does allow us to study the spatial variations
of the spectral properties in the X-ray emitting plasma.  In order to
analyze spectra extracted from physically uniform regions, we produced
a MOS mean photon energy map (i.e., an image where each pixel holds
the mean energy of the detected MOS photons) in the continuum band
$4.4-6.2$ keV. The shape and size of the seven regions were chosen in
order to have very small fluctuations of the mean photon energy
($\sim1\%$) and a sufficient number of photons.

Regions 1 and 2 are located on the X-ray eastern elongated structure
along the radio shell (labelled $a$/$\alpha$ in
Fig.~\ref{fig:1-9-radio}), respectively at its northern and southern
ends.  They both exhibit high values in the equivalent width maps.

Regions 3 and 4 cover the brightest part of the centrally barrel-shaped 
X-ray structure: region 3 corresponds to the maximum of the continuum 
emission in the $4.4-6.2$ keV band (Fig. \ref{fig:lines}, upper left panel), 
region 4 corresponds to the highest Fe K line emission region 
(Fig.~\ref{fig:lines}, lower left panel). 

Region 5 follows the western radio structure $\gamma$
(Fig.~\ref{fig:1-9-radio}).  Region 6 covers the western X-ray
structure $c$ and, finally, region 7 lies on the faint south-western
structure that is seen in the continuum and K line emission of S, Ar,
Ca, but not of iron.

\subsubsection{Spectral modelling and results}
\label{spec results}

We first modelled the extracted spectra using XSPEC with a single
MEKAL model of an optically-thin thermal plasma in collisional
ionization equilibrium (\citealt{mgv85}, \citealt{mlv86},
\citealt{log95}). Good fits to the spectra require non-solar
abundances for Si, S, Ar, Ca and Fe.  All other element abundances
were fixed at their solar values.  To account for the contribution
from the Cr line (Sect.~\ref{Global spectrum}), we added a narrow
gaussian component with fixed energy; it was not necessary to account
for the Mn line. We also added a systematic 5\% error term to reflect
the estimated 5-10\% uncertainties in the calibration of the
instrumental effective area.  The reduced $\chi^{2}$ values obtained
using the two MOS and the pn cameras range from 1.3 (with 354
d.o.f.) for region 5, to 1.7 (445 d.o.f.) for region 2. Above 2 keV,
the models describe the continuum and emission line spectra quite
well, but these single-temperature models severely underpredict the Si
XIII K$\alpha$ line, as illustrated in Fig.~\ref{fig:spettri2} (left
panel).  The best-fit parameters for these fits are thus not reported
here. The same problem was encountered by \citet{hph00} in the
modelling of the $ASCA$ global spectrum. We found that the quality of
the fits does not significantly improve if a single non-equilibrium
ionization (NEI) model is used instead.

We obtained a good description of the spectra by adding a second
collisional ionization equilibrium thermal MEKAL component. We require
the two components to have the same abundances and to be absorbed by
the same interstellar column density. The spectra are also
well-described if the soft component has solar abundances and is in
nonequilibrium ionization (PSHOCK model). In this case the
temperatures of the two components (NEI+CIE) are comparable to those
obtained for the two CIE components, but the fits give somewhat higher
chi-squared values. For example, in region 2:
$\chi^{2}_{NEI+CIE}=529.7$ (with 442 d. o. f.) and
$\chi^{2}_{2CIE}=506.7$ (443); in region 4: $\chi^{2}_{NEI+CIE}=821.2$
(673) and $\chi^{2}_{2CIE}=793.5$ (674); in region 6:
$\chi^{2}_{NEI+CIE}=1064.1$ (843) and $\chi^{2}_{2CIE}=1059.3$
(844). We therefore adopt the two CIE component model hereafter, while
cautioning the reader that the high interstellar absorption
complicates the description of the soft component and hence the
discrimination between these two models.

Fig.~\ref{fig:spettri2} shows a representative spectrum, extracted
from region 2, with best-fit model and residuals for two thermal
components (right panel), and for comparison, a single thermal
component (left panel). The best-fit parameters for all seven spectral
regions are given in Table~\ref{tab:bestfit}.

The values of $N_{\rm H}$ in the different spectral regions range
approximately from $4.2\times10^{22}$~cm$^{-2}$ to
$4.9\times10^{22}$~cm$^{-2}$ (compared with the global $N_{\rm
H}=5.0^{+0.3}_{-0.2}\times10^{22}$~cm$^{-2}$ estimated by
\citealt{hph00}).  To produce the line and the equivalent width images
(see Sect.~\ref{line images}), we estimate the continuum contribution
under the lines by assuming a uniform absorbed slope of the continuum
based on the global spectrum of W49B. For low energy lines, spatial
variation of the column density over the remnant can thus introduce
errors in the line and equivalent width images.  In particular, where
$N_{\rm H}$ is higher than its global averaged value, we may
overestimate the continuum under the line and hence underestimate the
equivalent width. We observe this kind of effect in the S equivalent
width image (Fig. \ref{fig:EQW}, upper left panel), where there is a
minimum in region 3 with a mean value of 0.32 keV, corresponding to
the highest value of the column ($N_{\rm H}=4.8^{+0.1}_{-0.2}$), and a
higher mean equivalent width of 0.4 keV in region 2 which has a lower
column ($N_{\rm H}=4.22^{+0.08}_{-0.05}$).

Variations in column density may also affect the Si images, but in
this case there is no a clear relation between such variations and
inhomogeneities in the equivalent width map. This probably indicates
that other effects are also present. As explained above, a single
temperature model provides a good overall description of the spectra
in the $1-9$ keV band, but fails to reproduce the observed Si line
flux in the narrow energy range $1.7-1.9$ keV. It is therefore
possible that the Si He-like emission originates predominantly in a
plasma with different physical conditions than that responsible for
the other emission lines. As shown in Fig. \ref{fig:Si}, the He-like
line morphology resembles that of the Si H-like emission as well as
the S, Ar and Ca line emission. The Si equivalent width maps (not
presented here) have irregular patchy features at odds both with those
seen for the other elements, and with the spatially resolved spectral
analysis. This is probably because the continuum of the hard component
is not negligible at low energies, even though the Si He-like line
emission itself is associated with the soft thermal component (see the
right panel of Fig. \ref{fig:spettri2}).

We observe larger variations for the temperature of the hotter
component $kT_{2}$ ($1.75-3.3$~keV) than for the cooler component
$kT_{1}$ ($0.7-1.05$~keV), but do not measure any radial temperature
structure (see Table~\ref{tab:bestfit}).  The temperature of the
hotter component is at a maximum in Region 1 at $kT_{2}$, and
decreases significantly moving southward along the X-ray elongated $a$
structure to region 2. This temperature also decreases towards the
western side of the remnant (from region 3 to region 6).  We observe
that the difference between the continuum surface brightness of
regions 3 (i. e. the brightest region in the continuum band $4.4-6.2$
keV) and region 4 is caused by the higher temperature in region 3, and
not by differences in the emission measure.

In Table~\ref{tab:bestfit}, the parameter $EM$ indicates the values of
the emission measure per unit area, which is the product of the square
of the particle density times the extension of the plasma along the
line of sight.  These values are in the range
$0.3-1.7\times~10^{20}$~cm$^{-5}$ for the cooler component, and
$0.4-3.5\times~10^{20}$~cm$^{-5}$ for the hotter component. In region
6 we have the highest values for both $EM_{1}$ and
$EM_{2}$. Associated with the decreasing trend of $T_{2}$, we note
that in region 6 the corresponding emission measure $EM_{2}$ is
slightly higher than in regions 3 and 4.

We used the best-fit values of temperature and emission measure to
derive the plasma density (indicated with $n$ in Table
\ref{tab:bestfit}) and the filling factor ($f$) of each component,
obtained as in \citet{bms99}. These quantities were derived assuming
pressure equilibrium between the two components and assuming that in
each spectral region the extension of the X-ray emitting plasma is
equal to the length of the chord intercepted by the radio shell
(approximated as a sphere with angular radius $\phi = 2.15'$) along
the line of sight. The filling factor of the cool component is $\la
0.1$, and the densities of the two components are almost uniform
across the remnant.

We found significant element overabundances in all the spectral
regions: all the elements have fitted abundances $>1.5$ (at 90$\%$
confidence level) with respect to their solar value. This attests to
the origin of the emission in the ejecta. Moreover, higher abundance
values are found for the central and eastern regions compared to the
western regions, in good agreement with the results of our image
analysis.  All abundances are lower in regions 5, 6, and 7, with the
largest decrease observed for the Fe abundance, as expected. For
example, the ratio $(Ar/Ar_{\odot})/(Fe/Fe_{\odot})$ is $0.7\pm0.1$ in
region 4 and $1.8\pm0.3$ in region 6, and the ratio
$(Ca/Ca_{\odot})/(Fe/Fe_{\odot})$ is $1.2\pm0.2$ in region 4 and
$1.8^{+0.3}_{-0.2}$ in region 6. This is in agreement with the
considerations discussed at the end of Sect.~\ref{line images}, where
regions 6 and 7 show low equivalent widths for iron, but nonnegligible
line emission and line equivalent widths for the lower-Z
elements. All the abundances reported in
Table~\ref{tab:bestfit} are given as the ratio of the number density
of the respective element to the number density of hydrogen expressed
in units of the solar value of this ratio. In a pure-ejecta plasma
(where hydrogen emission is negligible) this ratio would be
ill-defined, and the spectral shape would depend only on the relative
abundances of the heavy elements. We verified that this is not the
case for our fits.  In fact if we scale all the abundances by the same
factor $f$ we obtain significantly higher $\chi^{2}$ values for
$f\ga1.5$. Therefore the abundances in Table \ref{tab:bestfit} are
well defined and above solar.

In each spectral region we also obtained another $\chi^{2}$ minimum
with $kT_{1}\sim 0.2$ keV. This value is in agreement with the results
obtained with the $ASCA$ data by \citet{hph00} and \citet{kon05}. In
this case, however, we found very high values for the emission measure
per unit area of the cool component ($EM_{1}\ga10^{22}$ cm$^{-5}$),
which would imply unrealistically high values for the filling factor
($f_{1}\sim 0.5$) and the density ($n_{1}\sim 50$ cm$^{-3}$).  In what
follows, we will therefore discuss only the results presented in Table
\ref{tab:bestfit}.

\begin{center}
\begin{table*}[htb!]
\begin{center}
\caption{Best-fit parameters for the spectra extracted in the seven
regions shown in Fig. \ref{fig:lines} (lower left panel), described
with two thermal components in collisional ionization equilibrium. All
errors are at the 90\% confidence level.}
\begin{tabular}{lcccccccc} 
\hline\hline
   Parameter        &   Region   1       &    Region2          &       Region 3       &       Region 4      &   Region 5         &    Region 6       &   Region 7   & \\ \hline
$N_{\rm H}$ ($10^{22}$ cm$^{-2}$)& $4.5\pm0.1$ & $4.22^{+0.08}_{-0.05}$ & $4.8^{+0.1}_{-0.2}$ &     $4.5\pm0.1$     & $4.3^{+0.2}_{-0.1}$ &  $4.61\pm0.05$  & $4.36^{+0.08}_{-0.09}$\\
    $kT_{1}$ (keV)  &$0.90^{+0.08}_{-0.06}$&$0.75^{+0.03}_{-0.05}$&$0.92^{+0.06}_{-0.08}$& $0.8\pm0.1$ &  $0.96^{+0.09}_{-0.09}$ &$0.78^{+0.06}_{-0.04}$&$0.82^{+0.09}_{-0.06}$\\
$EM^{*}_{1}$ ($10^{20}$ cm$^{-5}$)&$0.5^{+0.2}_{-0.1}$&$1.16^{+0.1}_{-0.09}$&$1.2\pm0.2$&  $1.3^{+0.3}_{-0.3}$&$0.8^{+0.3}_{-0.2}$&$1.1^{+0.4}_{-0.2}$&$0.92^{+0.1}_{-0.05}$ \\
    $kT_{2}$ (keV) &$3.0^{+0.4}_{-0.2}$ &$2.19^{+0.1}_{-0.06}$ & $2.6\pm0.2$  & $2.35^{+0.12}_{-0.09}$ & $2.2^{+0.2}_{-0.2}$ &$1.77^{+0.03}_{-0.09}$ & $2.00\pm0.04$ \\
$EM^{*}_{2}$($10^{20}$ cm$^{-5}$)&  $0.5\pm0.1$  &  $1.1\pm0.1$  & $1.8^{+0.1}_{-0.1}$ &$1.95^{+0.1}_{-0.08}$&  $1.4\pm0.3$  & $2.2^{+0.3}_{-0.1}$  & $1.3\pm0.1$ \\
    $Si/Si_\odot$   & $2.2\pm0.4$ & $3.4^{+0.2}_{-0.4}$   & $3.7\pm0.4$  &  $2.9\pm0.3$  & $2.4^{+0.3}_{-0.5}$ & $2.3\pm0.1$  & $2.1^{+0.3}_{-0.2}$  \\
      $S/S_\odot$   &  $4.1\pm0.5$   &  $4.1\pm0.3$    & $4.6^{+0.3}_{-0.6}$ & $3.6^{+0.3}_{-0.4}$ & $2.7^{+0.3}_{-0.5}$  & $2.4^{+0.2}_{-0.1}$  & $2.8\pm0.3$ \\
    $Ar/Ar_\odot$   & $5.6^{+1.1}_{-0.9}$  & $3.8\pm0.6$   & $4.9^{+0.8}_{-0.6}$ & $4.1^{+0.3}_{-0.7}$ & $2.6\pm0.6$ & $2.8\pm0.3$  & $3.4^{+0.3}_{-0.5}$ \\
    $Ca/Ca_\odot$   &   $8\pm1$    & $6.6^{+0.9}_{-0.8}$  & $7\pm1$ & $6.9^{+0.7}_{-0.9}$  &  $4.7^{+0.8}_{-0.9}$ & $4.4^{+0.4}_{-0.5}$ & $5.2^{+0.5}_{-0.7}$ \\    
    $Fe/Fe_\odot$   & $5.2_{-0.4}^{+0.6}$  &  $5.1\pm0.3$   & $5.1^{+0.4}_{-0.6}$ & $5.6^{+0.4}_{-0.3}$ & $2.6^{+0.2}_{-0.3}$ & $2.5^{+0.1}_{-0.2}$ & $1.6^{+0.2}_{-0.1}$ \\    
 $\chi^{2}/$d. o. f.&  $298.8/244$    &   $506.7/443$     &  $800.0/646$    &   $793.5/674$   &   $403.2/352$   &  $1059.3/844$    &  $566.3/434$    \\    
\hline
 $f_{1}$   & $0.08^{+0.03}_{-0.04}$ &  $0.11^{+0.04}_{-0.03}$& $0.08 \pm 0.04$ & $0.08 \pm 0.03$ & $0.10^{+0.06}_{-0.05}$ & $0.09^{+0.06}_{-0.04}$ & $0.11^{+0.04}_{-0.03}$ \\
 $n_{1}$ (cm$^{-3}$)& $6^{+3}_{-2}$ & $7^{+2}_{-1}$ & $7^{+4}_{-2}$ & $8 \pm 2$ & $5^{+3}_{-2}$ & $7^{+4}_{-2}$ & $5.5^{+3}_{-0.9}$ \\
 $n_{2}$ (cm$^{-3}$)& $1.8^{+0.3}_{-0.2}$ & $2.5^{+0.2}_{-0.1}$ & $2.6\pm 0.2$ & $2.6^{+0.3}_{-0.1}$ & $2.3 \pm 0.2$ & $3.0^{+0.8}_{-0.2}$ & $2.3 \pm 0.2$ \\
\hline\hline
\multicolumn{8}{l}{\footnotesize{* Emission measure per unit area.}} \\
\label{tab:bestfit}
\end{tabular}
\end{center}
\end{table*}
\end{center}

\subsection{Evidence for Ni overabundance}
\label{Ni+Cr}

Thanks to the high \emph{XMM-Newton} effective area at high energies
it was possible to detect a strong emission feature at $\sim7.8$~keV,
with possible contributions from both Fe XXV and Ni XXVII. The
Ni XXVII emission line has been previously detected in only in one
SNR, Cas A (\citealt{bwv01}, \citealt{wbv02}). This emission blend
can be used to set constraints on the Ni abundance in W49B. Since the
photon statistics for the seven individual spectra do not allow
us to resolve the features in the $7.65-8.0$~keV
band in detail, we extracted the spectrum from the larger elliptical
region shown in Fig.~\ref{fig:lines} (lower right panel). This region
covers the area of highest surface brightness in the $7.65-8.0$ keV
band. We fitted the spectrum in the $1-9$ keV band with the
2-component thermal model used for the spatially resolved spectral
analysis.  Table~\ref{tab:Ni} gives the best-fit parameters derived
both by fixing the Ni abundance to the solar value (second column) and
by fitting the Ni abundance as a free parameter (third column).  We
can better reproduce the Fe XXV and Ni XXVII K line blends by
fitting for the Ni abundance.  The F-test gives the probability
that the improvement of the fit is insignificant as $<10^{-10}$. We
conclude that there is convincing evidence for Ni overabundance
($Ni/Ni_{\odot}=10^{+2}_{-1}$) in the centrally barrel-shaped
structure of W49B.
\begin{center}
\begin{table}[htb!]
\begin{center}
\caption{Best-fit parameters for the spectrum extracted from the
elliptical region shown in Fig.~\ref{fig:lines} (lower right panel)
and reproduced by two thermal components in collisional ionization
equilibrium. All errors are at the 90\% confidence level.}
\begin{tabular}{lccc} 
\hline\hline
        Parameter                  &   $Ni/Ni_{\odot}=1$    &     $Ni$ thawed        \\ \hline
$N_{\rm H}$ ($10^{22}$ cm$^{-2}$)  & $4.60^{+0.06}_{-0.03}$ & $4.68^{+0.06}_{-0.02}$ \\     
 $kT_{1}$ (keV)                    & $0.87^{+0.04}_{-0.02}$ & $0.86^{+0.02}_{-0.05}$ \\  
$EM^{*}_{1}$ ($10^{20}$ cm$^{-5}$) & $1.13^{+0.04}_{-0.02}$ & $1.1^{+0.1}_{-0.2}$ \\   
 $kT_{2}$ (keV)                    & $2.35^{+0.02}_{-0.03}$ &   $2.2^{+0.2}_{-0.1}$  \\ 
$EM^{*}_{2}$ ($10^{20}$ cm$^{-5}$) & $1.79^{+0.06}_{-0.04}$ & $1.82^{+0.06}_{-0.05}$ \\      
 $Si/Si_\odot$                     &  $3.3^{+0.1}_{-0.3}$   &     $3.3\pm{0.2}$     \\
 $S/S_\odot$                       &      $3.7\pm0.1$       &   $3.7^{+0.1}_{-0.2}$  \\
 $Ar/Ar_\odot$                     &  $4.2^{+0.4}_{-0.2}$   &	$4.2^{+0.3}_{-0.4}$  \\
 $Ca/Ca_\odot$                     &      $6.3\pm0.5$       &	  $6.4\pm{0.5}$     \\
 $Fe/Fe_\odot$                     &  $5.5^{+0.1}_{-0.2}$   &  $6.0^{+0.1}_{-0.2}$   \\
 $Ni/Ni_\odot$                     &            $1$    	    &      $10^{+4}_{-1}$     \\
 $\chi^{2}/$d. o. f.               &   $1847.8/1327$        &	   $1584.3/1326$     \\

\hline\hline
\multicolumn{3}{l}{\footnotesize{* Emission measure per unit area.}} \\
\label{tab:Ni}
\end{tabular}
\end{center}
\end{table}
\end{center}

\subsection{Overionization}
Through their analysis of the $ASCA$ global spectrum, \citet{kon05}
claimed the presence of ``overionized'' plasma in W49B. In particular,
they measured the intensity ratio of the H-like to the He-like
K$\alpha$ lines of Ar and Ca to obtain the ionization temperature
$T_{z}$, and found that $T_{z}$ is higher than the electron
temperature $T_{e}$ estimated from the continuum of the X-ray
spectrum. We performed the same analysis by modelling the
\emph{XMM-Newton} global spectrum in the $2.75-6.0$ keV band.  We
model the continuum and estimate $T_{e}$ with a zero metallicity
VMEKAL component and include four narrow gaussian components to model
the Ar and Ca lines and a fifth gaussian component for the Cr line. We
considered absorption column densities of $N_{H}=4.6\times 10^{22}$
cm$^{-2}$ and $N_{H}=5.23\times 10^{22}$ cm$^{-2}$ (the value adopted
by \citet{kon05}). For those column densities, we obtained electron
temperatures of $kT_{e}=1.80\pm0.03$ keV and $1.66^{+0.02}_{-0.01}$
keV, respectively.  We find no evidence of overionization for Ar,
given the respective ionization temperatures of $kT_{z}=1.68\pm0.07$
and $kT_{z}=1.64^{+0.08}_{-0.05}$.  For Ca, the ionization temperature
does seem higher than the continuum temperature:
$kT_{z}=2.0^{+0.1}_{-0.05}$, with $N_{H}=4.6\times 10^{22}$ cm$^{-2}$,
and $kT_{z}=2.04^{+0.1}_{-0.08}$, with $N_{H}=5.23\times 10^{22}$
cm$^{-2}$. However, we caution the reader that these estimates may not
be reliable because the global spectrum originates from physically
non-uniform regions of W49B that have different temperatures and
abundances.  A case in point in that the estimates of the electron
temperatures obtained from the global spectrum are significantly lower
those obtained from our spatially resolved spectral analysis of
homogeneous regions (see Sect. \ref{spectral analysis} and $kT_{2}$
values in Table \ref{tab:bestfit}). We therefore repeated our analysis
for the more uniform central region indicated by a white ellipse in
the lower left panel of Fig. \ref{fig:EQW}.  This is where the
overionization effects are expected to be strongest in the scenario
proposed by \citet{kon05}. Our analysis shows that these effects are
not present. Using the same model described above in the 2.75-6.0 keV
band, we found for $N_{H}=4.6\times 10^{22}$ cm$^{-2}$ and
$N_{H}=5.23\times 10^{22}$ cm$^{-2}$: respective electron temperatures
of $kT_{e}=2.4\pm0.1$ keV and $kT_{e}=2.25^{+0.08}_{-0.1}$, Ar
ionization temperatures of $kT_{z}=1.9^{+0.1}_{-0.2}$ and
$kT_{z}=1.8\pm{0.1}$, and Ca ionization temperatures of
$kT_{z}=2.3^{+0.1}_{-0.3}$ and $kT_{z}=2.2\pm{0.2}$.  Given that
$kT_{z}$ is always slightly lower than, or consistent with $kT_{e}$, we
conclude that there is no indication of overionization in the central
region of W49B.

\section{Discussion}
\label{Discussion}

On the basis of low angular resolution $ASCA$ line images,
\citet{fti95} argued for stratification of the SN ejecta in W49B, with
iron confined to the inner region and lower-Z elements (Si and S)
forming an outer envelope.  This hypothesis was supported by the
analysis of the global spectrum, with the intensity ratios of
Ly$\alpha$ and He$\alpha$ giving different ionization ages for Si, S,
Ar and Ca.  \citet{hph00} pointed out that while this conclusion is
compatible with the $ASCA$ spectrum, no difference in the ionization
ages is required if the data are modelled with a two-component thermal
plasma.  The analysis of \emph{XMM-Newton} data allowed us to
discriminate between these two scenarios. Our spatially resolved
spectral analysis has shown that a correct fit of the spectra requires
two thermal components and that these two components are compatible
with collisional ionization equilibrium, both at the center and in the
outer parts of the remnant. Moreover, as explained in Sect.~\ref{line
images}, our line images and equivalent width maps exclude a radial
stratification of the elements in the ejecta, with Fe K emission
confined to the inner part of the X-ray emitting region and lower-Z
emission lines originating in an outer shell.  In fact, both the
central barrel-shaped structure and the eastern elongated structure of
the SNR have high abundances of Ar, Ca and Fe, while the western part
has lower abundances.

The physical conditions of the plasma are not uniform in W49B. The
highest temperatures are found in the northern end of the elongated
eastern structure (labelled $a$) and in the central barrel-shaped
structure (labelled $b$). The lowest temperatures are found in the
western part of W49B and in the south-eastern zone. Assuming pressure
equilibrium between the two components, no significant variations in
the plasma density across the remnant are observed.  The observed
X-ray emission originates in the SN ejecta, and Si, S, Ar, Ca, Fe and
Ni are all highly overabundant with respect to their solar values.
They have almost the same values in structures $a$ and $b$, and are
lower (but still overabundant) in structure $c$. No clear evidence for
X-ray emission from the shocked ambient medium is present. In fact,
the results of the spatially resolved spectral analysis suggest that
the both the X-ray emitting components are associated with
overabundant plasma.  This is in agreement with the similarity of the
morphology between the Si He-like emission blend (associated with the
cooler component) and the other emission blends (associated with the
hotter component).  As explained in Sect. \ref{spectral analysis},
however, the high absorption column inhibits the possibility of a
detailed study of the soft component in that it is not easy to
determine its morphology. Furthermore, the contribution of the hot
component is significant even below 1.7 keV, so that it is not
possible to produce a clean image of the soft component. We therefore
cannot rule out the possibility that the soft component may be
associated with the shocked circumstellar or interstellar medium.

It is interesting to consider to what extent the complex morphology of
W49B, and the inhomogeneities in temperature and abundance, are caused
by the dynamics of the SN explosion or by the remnant's interaction
with its environment.  Recent 2.12 micron infrared imaging
observations have revealed a shocked molecular H$_{2}$ cloud that
confines the remnant (Fig. \ref{fig:chandraIR}). The circular
distribution of the molecular cloud around the SNR has likely been
shaped by strong stellar winds from the progenitor star pushing out
the ambient interstellar material to its current location and forming
a cavity.  Infrared [Fe II] observations at 1.64 $\mu m$ reveal a
barrel-shaped structure with four coaxial rings of warm gas, which has
been interpreted as residuals of a strong stellar wind
(\citealt{krr04}).

The morphology of the eastern region of W49B has clearly been shaped
by the ambient molecular cloud.  Despite their similar abundance
values, the eastern $a$ and central $b$ regions have very different
morphologies. The fact that region $a$ is almost perpendicular to
region $b$ is probably due to its confinement by the large H$_{2}$
cloud to the east.  The [Fe II] observations show that the cavity is
an inhomogeneous environment for the supernova remnant expansion.

In contrast, there are no observed interstellar structures that might
have shaped the elongated structure of the central X-ray barrel $b$.
This centrally barrel-shaped structure thus may have been produced
directly by the supernova explosion or by pre-existing structures in
the ambient stellar wind material.

The central barrel morphology seen in the \emph{Chandra} Fe K image
has been attributed to a bipolar jet, and taken as evidence that W49B
is a $\gamma-$ray burst (GRB) remnant\footnote
{http://chandra.harvard.edu/press/04$\_$releases/press$\_$060204.html}.
The presence of such an elongated morphology, however, does not
necessarily reflect a jet-like explosion. A similar elongated central
bright structure has been observed with $Chandra$ in the G292.08$+$1.8
remnant, but \citet{phs04} showed that its emission is associated with
normal chemical composition plasma that originates in shocked dense
circumstellar material rather than shocked ejecta.  Unlike for
G292.08$+$1.8 SNR, our analysis suggests that the X-ray emission in
the central X-ray barrel of W49B is dominated by ejecta, although we
cannot rule out the possibility that the soft component comes from
solar abundance material.  The scenario proposed for G292.08$+$1.8 can
nevertheless be invoked for W49B. It is possible that the barrel-like
X-ray emission reflects the morphology of the ambient medium across
the remnant. While no dense structures are visible toward the North
and South of W49B (as shown in the infrared image of
Fig. \ref{fig:chandraIR}), we may assume the ambient medium to be
enhanced in the central zone, forming a torus-like structure all
across W49B. According to this scenario, the cold component may be
associated with the thin belt of material heated by the transmitted
shock travelling into the ISM structure. This would explain the low
values of the filling factor of the soft component. The hot component
is associated with the ejecta interacting with the reflected shock. In
this case, only the material expelled in the plane of the torus is interacting with the the reverse shock and heated by it. Notice, however, that according to our estimates
of the plasma density for the spectral regions, the total mass of the
ejecta in the hot component in these regions is $\sim 8$
M$_{\odot}$. This indicates that if we assume that we are observing
only a small fraction of the ejecta we probably overestimate the total
expelled mass.
 
Another possibility is that the barrel is associated with a bipolar
ejecta jet interacting with the cavity wall visible at the eastern end
of the remnant. In this case we have to assume that the density
contrast has driven a very strong reverse shock which has now reached
the center of W49B.  Moreover, we have found that structures $a$ and $b$
have similar abundance values and therefore, that $a$ could be considered
as the head of the eastern jet. The scenario that emerges is that this
jet has been distorted to deviate mostly southwards by its impact with
the H$_{2}$ wall. In this framework, we would have a highly
anisotropic bipolar jet, where the eastern arm is significantly more
extended, hot and Fe-rich than the western one.

Jet-like explosions have been investigated as possible mechanisms for
$\gamma-$ray bursts (GRB) because a spherically symmetric GRB would
require an extremely large energy (several times the solar rest
mass). A link between very energetic SNe and GRBs has been found by
\citet{gvv98}, who showed that the bright type Ic SN 1998bw (with an
estimated energy $E\sim3\times10^{52}$~erg, \citealt{imn98}) is the
probable optical counterpart of the $\gamma-$ray burst GRB
980425. Observations of SN 1998bw have shown strong indications of
asphericity in the explosion (\citealt{mnp01}). Indications of
aspherical explosions have also been found in many other hypernovae,
i. e. supernova explosions with $E\ga10^{52}$~erg (for a review on
hypernovae see \citealt{nmu03}).

The relationship between asymmetric hypernovae and GRB is in agreement
with the collapsar model (\citealt{mw99}), in which the collapse of a
massive rotating core can generate a black hole and supersonic jets
which propagate through the star. The nebular spectrum of SN 1998bw is
well reproduced by such a model. \citet{kho99} have shown that this
mechanism produces a highly aspherical explosion with bipolar fast
jets. A bipolar explosion does not necessarily imply a GRB, however,
because a high explosion energy ($\ga10^{52}$ erg) is also required.

To investigate the possibility that W49B is a GRB remnant, we compared
the abundance values found in the $a$ barrel (see Table~\ref{tab:Ni})
with those obtained from models of explosive nucleosynthesis in
bipolar explosions by \citet{mn03}. In particular, we compared the
expected and observed ratios $(X/X_{\odot})/(Fe/Fe_{\odot})$ for
$X=Si$, $S$, $Ar$, $Ca$, $Cr$, $Mn$ and $Ni$ (for Cr and Mn we assumed
the same ratios as observed for Ar, in agreement with the estimates
given in \citealt{hph00}). None of the models can reproduce the entire
observed abundance pattern. The hypernova models with zero age main
sequence progenitor mass $M_{ZAMS}=40\rm M_{\odot}$ and explosion
energy $E>10^{52}$ erg are strongly in disagreement with the observed
values (with very large discrepancies for the Si, S and Ar
abundances).  We obtained better results with the less energetic
models 25A ($E=6.7\times 10^{51}$ and $M_{ZAMS}=25\rm M_{\odot}$) and
25B ($E=0.6\times 10^{51}$ and $M_{ZAMS}=25\rm M_{\odot}$). We also
found relatively good agreement with the spherical explosion model
25Sa ($E=1\times 10^{51}$ and $M_{ZAMS}=25\rm M_{\odot}$). The
predicted values of these three models and our observational findings
are summarized in Fig.~\ref{fig:abund}.  There is good overall
agreement for Ar through Ni, but the observed abundances of Si and S
are not well reproduced.

While the observed abundance pattern in W49B does not favor an
hypernova explosion of sufficient energy to generate a GRB, we cannot
definitely exclude that W49B was related to such an event.

\section{Summary and conclusions}
\label{Summary and conclusions}

Our study of the \emph{XMM-Newton} observations of the Galactic supernova 
remnant W49B has shown that:
\begin{itemize} 
\item The remnant presents a complex X-ray morphology characterized by
centrally barrel-shaped emission ($b$), terminated on the eastern side
by an elongated perpendicular structure ($a$) and on the western side
by a more diffuse, nearly aligned structure ($c$).
\item The physical conditions of the plasma are not homogeneous
throughout the remnant. Lower temperatures are found in remnant's
western end, while hotter regions are located in the center and
northeast.
\item A high Ni overabundance ($Ni/Ni_{\odot}=10^{+2}_{-1}$) has been
measured in the central barrel, and the presence of Cr and Mn lines in
the global spectrum is confirmed.
\item We found no evidence for overionization of the plasma in the
center of W49B.
\item The X-ray emission from all the observed regions of W49B
originates in the ejecta, based on significant element overabundances
of Si, S, Ar, Ca and Fe. The equivalent width maps of W49B and a
detailed spectral analysis of individual regions both show that the
central ($b$) and eastern ($a$) structures have similar abundances,
and the western ($c$) region exhibits lower abundances. This decrease
is particularly pronounced for iron.
\item The X-ray morphology of W49B may be attributed to pre-existing
structures in the ambient medium which may have produced an aspherical
reverse shock, or alternatively to an aspherical supernova explosion
in a wind cavity (bounded by molecular clouds on the eastern side of
the remnant). Because it has abundances similar to those in the
central barrel, the eastern structure $a$ can be considered as the
head of the jet.  It is distorted and deviates mostly southward due to
its impact with the dense molecular wall that confines its expansion
to the east.
\item We investigated the possibility that W49B is a $\gamma$-ray
burst remnant by comparing the observed abundances with yields derived
from supernova and hypernova nucleosynthesis models. We found better
agreement with the less energetic ($E\sim10^{51}$~erg) models.  The
association of W49B with a GRB is therefore not necessarily implied by
the spectral data.
\end{itemize}

\begin{acknowledgements}

We thank the referee, Dr B. Aschenbach, for his comments and
suggestions.  The authors wish to thank Joseph Lazio for providing us
with the 327 MHz radio image of W49B. M. M. thanks Costanza Argiroffi
for the discussions about the ionization temperature. This work was
partially supported by the Minist\`{e}re fran\c{c}ais des Affaires
Etrang\`{e}res.

\end{acknowledgements}

\bibliographystyle{aa}
\bibliography{references}

\begin{figure}[h!]
\caption{\emph{Left panel:} Vignetting corrected EPIC count rate 
image of W49B in the $1-9$ keV band. The image is adaptively smoothed (with 
signal-to-noise ratio 10) and background-subtracted. The 2 pc scale has been 
obtained assuming a distance of 8 kpc. The count rate ranges between 0 and 
$7.2\times 10^{-3}$ s$^{-1}$. \emph{Right panel:} Radio image of W49B at 327 
MHz (angular resolution $6''$), obtained by \citet{llk01}. We have 
superimposed in black the X-ray contour levels derived from the $1-9$ keV 
EPIC image at $25\%$, $50\%$ and $75\%$ of the maximum. The color bar has a linear scale.}
\label{fig:1-9-radio}
\end{figure}

\begin{figure}[h!]
\caption{Color composite \emph{Chandra} X-ray image (in blue), infrared 2.12 $\mu$m molecular hydrogen (in red), and 1.64 $\mu$m 
[Fe II] (in green) images of W49B (from http://chandra.harvard.edu/press/04$\_$releases/press$\_$060204.html).}
\label{fig:chandraIR}
\end{figure}

\begin{figure}[h!]
\caption{pn (upper) and MOS (lower) global spectra of W49B in the
$1-9$ keV energy band, extracted from a circular region covering the
entire remnant (of radius $3.3'$ corresponding to $\sim 7.7$ pc).
Line energies have been determined using narrow gaussian components on
top of a thermal bremsstrahlung continuum. Note that to reproduce the
continuum at energies below the Si K lines, an additional
bremsstrahlung component is required.}
\label{fig:specTOT}
\end{figure}

\begin{figure}[h!]
\caption{Vignetting corrected and continuum-subtracted EPIC count rate
images of the continuum ($4.4-6.2$ keV, \emph{upper left}); S XV K
($2.35-2.53$ keV, \emph{upper right}); Ar XVII K ($3.05-3.2$ keV,
\emph{central left}); Ca XIX K ($3.8-4.0$ keV, \emph{central right});
Fe XXV K ($6.45-6.9$ keV, \emph{lower left}) and Fe XXV$+$Ni XXVII K
($7.65-8.0$ keV, \emph{lower right}). All images are adaptively
smoothed with signal-to-noise ratio 10. The 7 regions selected for
spatially resolved spectral analysis are shown in the Fe XXV K line
image, while the ellipse in the Fe XXV$+$Ni XXVII K line image
indicates the region used for the detection of the Ni line
(Sect. \ref{Ni+Cr}). The count rates range between 0 and:
$8.6\times10^{-4}$ s$^{-1}$ (continuum); $3.4\times10^{-4}$ s$^{-1}$
(S); $1.7\times10^{-4}$ s$^{-1}$ (Ar); $1.9\times10^{-4}$ s$^{-1}$
(Ca); $7.0\times10^{-4}$ s$^{-1}$ (Fe); $2.1\times10^{-5}$ s$^{-1}$
(Fe$+$Ni). The color bar has a linear scale.}
\label{fig:lines}
\end{figure}

\begin{figure}[h!]
\caption{Equivalent width maps of the S He-like and H-like line
emission in the $2.35-2.7$ keV band (\emph{upper left}); of the Ar
XVII K blend ($3.05-3.2$ keV, \emph{upper right}); of Ca XIX K
($2.35-2.7$ keV, \emph{central left}); of Fe XXV K ($6.45-6.9$ keV,
\emph{central right}) and of the Fe XXV$+$Ni XXVII K emission lines in
the $7.65-8.0$ keV band (\emph{lower left}). In the \emph{lower right}
panel we have overlaid on the Fe XXV K equivalent width map the
$75\%$, $50\%$ and $25\%$ contour levels of the corresponding line
image (Fig. \ref{fig:lines}, lower left panel). The ranges of the
equivalent width maps are: $0.245-0.6$ keV (S); $0.07-0.13$ keV (Ar);
$0.085-0.18$ keV (Ca); $0-12.1$ keV (Fe); $0.315-1.4$ keV (Fe$+$Ni).}
\label{fig:EQW}
\end{figure}

\begin{figure}[h!]
\caption{Vignetting corrected and continuum-subtracted EPIC count rate
image of the Si XIII K$\alpha$ ($1.77-1.9$ keV, \emph{left panel}) and
Si XIV K$\alpha$ ($1.96-2.06$ keV, \emph{right panel}). The images are
adaptively smoothed with signal-to-noise ratio 10. The count rates
range between 0 and $2.3\times10^{-4}$ s$^{-1}$ for Si XIII and
$1.7\times10^{-4}$ s$^{-1}$ for Si XIV. The color bar has a
linear scale.}
\label{fig:Si}
\end{figure}

\begin{figure}[h!]
\caption{pn (upper) and MOS (lower) spectra of region 2 with the
corresponding best-fit model and residuals obtained with one thermal
component (left panel) and with two thermal components (right
panel). The contribution of the two components for the MOS camera is
indicated in the right panel in cyan for the cooler component and blue
for the hotter component.  }
\label{fig:spettri2}
\end{figure}

\begin{figure}[h!]
  \caption{Elemental abundances in the central X-ray barrel of W49B (red
crosses, see Table \ref{tab:Ni}), normalized to the Fe abundance
relative to the solar values (\citealt{ag89}), compared with the
yields obtained from the aspherical explosion models 25a (boxes) and
25b (diamonds) and from the spherical model 25Sa (stars) of explosive
nucleosynthesis by \citet{mn03}. The Cr and Mn abundances in
the central barrel have been derived from the Ar abundance following
\citet{hph00}.}
\label{fig:abund}
\end{figure}

\end{document}